\documentclass[pre,aps,eqsecnum]{revtex4}
\usepackage{epsfig}
\usepackage{amssymb}

\begin{document}

\title{Quantum Kramers turnover: a phase space function approach}
\author{Debashis~Barik and Deb~Shankar~Ray{\footnote{Email Address:
pcdsr@mahendra.iacs.res.in}}} \affiliation{Indian Association for
the Cultivation of Science, Jadavpur, Kolkata 700 032, India}

\begin{abstract}
The problem of Kramers' turnover is a central issue of dynamical
theory of reaction rate. Since its classical solution in the
Markovian limit in mid-eighties by Melnikov and Meshkov , the
problem has been addressed by a number of groups in the last
decade both in classical non-Markovian and quantum mechanical
context. Based on a coherent state representation of noise
operators and a positive definite Wigner canonical thermal
distribution function we have recently developed a c-number
quantum Langevin equation [Barik \textit{et al}, J. Chem. Phys.
{\bf 119}, 680 (2003); Banerjee \textit{et al}, Phys. Rev. E {\bf
65}, 021109 (2002)]. We implement this scheme within Pollak's
well known normal mode description to calculate the quantum
transmission coefficient over an arbitrary range of friction,
noise correlation and temperature. The theory generalizes the
quantum correction to Grote-Hynes factor in the rate expression
down to vacuum limit which reduces to well known high temperature
quantum correction, \textit{i.e.}, the Wolynes term for quantum
transmission and reflection for the barrier in the appropriate
limit and also considers the quantum corrections due to
nonlinearity of the system potential order by order which
contributes to energy loss and dispersion due to coupling between
unstable and stable normal modes near the barrier top and is
valid for both above and below the activated tunneling regime.
Our results have been compared with those obtained earlier for a
model potential and found to be good agreement.
\end{abstract}

\maketitle

\section{{\bf Introduction}}
Kramers' \cite{k1} diffusion model of chemical reactions proposed
in 1940 forms the dynamical basis of modern rate theory of
activated processes. The seminal and essential content of this
nonequilibrium formulation is the inclusion of dependence of rate
constant of a reaction on viscosity or friction of the reaction
medium. Based on the classical theory of Brownian motion in phase
space Kramers derived the expressions for nonequilibrium steady
state distribution functions to work out the rate coefficients in
the two different limiting situations and showed that the rate
varies linearly in the weak dissipation regime and inversely in
the high dissipation regime with friction. That is, in between the
energy diffusion and spatial diffusion limited regimes the rate
constant as a function of friction exhibits a bell-shaped curve -
known as Kramers' turnover. With the advent of ultrafast lasers
and time-resolved detection techniques since late seventies,
experimental confirmation of Kramers' theory provided a new
impetus for further development in chemical dynamics and
condensed matter physics \cite{dn,cs,lut,man,ot}.

In spite of this spectacular growth one point however, that
remained illusive for several decades was the absence of an
analytic rate formulae which correctly interpolates the two
limits. A significant advancement was made in 1986 when Melnikov
and Meshkov \cite{mel,mel1} proposed a theory of Kramers turnover
for Markovian friction. The key point of this analysis is the
calculation of an average energy loss of the system due to its
weak coupling with the bath modes near the top of the potential
barrier. The approach was further extended to non-Markovian
domain by Pollak, Grabert and H\"{a}nggi \cite{pgh} within the
framework of a normal mode description of the system plus
reservoir, where the implementation of an \textit{ad hoc} function
in the expression for the spatial diffusion limited rate could be
avoided. Subsequent to this formulation of classical theory, the
problem of quantum Kramers' turnover was addressed by Rips and
Pollak \cite{rip}. The degree of accuracy of classical and quantum
theories has been tested by numerical simulation of reaction rate
with model potentials. We refer to Refs.11-16 for further details.

The classical theory \cite{pgh} of Kramers' turnover uses a
Langevin description that governs the dynamics of the system by
an infinite number of harmonic oscillators coupled linearly to
the system degree of freedom. Very recently based on a coherent
state representation of noise operator and a positive definite
Wigner canonical thermal distribution of harmonic oscillators
\cite{hil} of the bath we have proposed a c-number quantum
Langevin equation \cite{db1,skb,db2,db3,bk1,bk2} in the context
of quantum rate theory and stochastic processes in terms of phase
space function formalism \cite{tan,tan1}. It would seem that one
should be able to analyze a quantum turnover for arbitrary noise
correlation and temperature by using such a c-number formulation.
This is the main purpose of this paper. In what follows we make
use of a c-number Hamiltonian in the normal mode procedure and
take care of the nonequilibrium dynamics at the barrier top by
calculating the average quantum energy loss of the unstable mode
due to its weak coupling with the stable bath modes and
equilibriation in the well by a Wigner thermal distribution.
Besides being an approach based on canonical quantization the
theory specifically addresses the following three points: First,
we quantize the system mode to make the theory applicable beyond
the activated tunneling regime down to absolute zero. Second,
although harmonic approximation is a good description at the
barrier top, actual shape and nonlinearity of the system
potential is important at low temperatures; but systematic
corrections to harmonic approximation have hardly been envisaged.
The present formulation takes care of quantum corrections due to
nonlinearity of the system degrees of freedom order by order.
Again even in the lowest order this quantum correction can be, in
principle, a contribution in the weak coupling between the stable
and unstable modes for calculation of energy loss and dispersion
as shown in this paper. Third, we take into account of the
quantum correction to Grote-Hynes dynamical factor in the rate
expression down to vaccum limit. This reduces to well-known high
temperature quantum correction or the Wolynes \cite{wolyn} term in
the appropriate limit. The approach thus generalizes the Wolynes
term in the deep tunneling regime.

The outlay of the paper is as follows. In the following section
II we discuss our c-number quantum Langevin equation which allows
us to realize a c-number Hamiltonian analyzed by Pollak's normal
mode analysis. The equilibrium theory in terms of Wigner
distribution function to formulate a quantum counterpart of
multidimensional transition state theory (TST) has been presented
in Sec.III. Sec.IV is devoted to nonequilibrium dynamics of the
unstable mode and its average energy loss over round trip time in
the well near the barrier top. Since this energy is sensitive to
quantum contribution due to nonlinearity of the potential we
calculate the quantum correction to energy loss of the unstable
mode explicitly in Sec.V. An explicit example with cubic
potential has been worked out in Sec.VI to compare with the known
results. The paper is concluded in Sec.VII.

\section{{\bf C-number quantum Langevin equation}}

We consider a particle of unit mass coupled to a medium comprised
of a set of harmonic oscillators with frequency $\omega_i$. This
is described by the following Hamiltonian:

\begin{equation}\label{1}
H = \frac{\hat{p}^2}{2} + V(\hat{q}) + \sum^N_{i=1} \left \{
\frac{\hat{p}_i^2}{2} + \frac{1}{2} (\omega_i \hat{x}_i -
\frac{c_i}{\omega_i} \hat{q} )^2 \right \}
\end{equation}

Here $\hat{q}$ and $\hat{p}$ are co-ordinate and momentum
operators of the particle and the set $\{ \hat{x}_i,\hat{p}_i \}$
is the set of co-ordinate and momentum operators for the
reservoir oscillators coupled linearly to the system through
their coupling coefficients $c_i$. The potential $V(\hat{q})$ is
due to the external force field for the Brownian particle. The
co-ordinate and momentum operators follow the usual commutation
relations [$\hat{q}, \hat{p}$] = $i \hbar$ and [$\hat{x}_i,
\hat{p}_j$] = $i \hbar \delta_{ij}$.

Eliminating the reservoir degrees of freedom in the usual way we
obtain the operator Langevin equation for the particle,

\begin{equation}\label{2}
\ddot{ \hat{q} } (t) + \int_0^t dt'  \gamma (t-t') \dot{ \hat{q}
} (t') + V' ( \hat{q} ) = \hat{F} (t) \; \; ,
\end{equation}

where the noise operator $\hat{F} (t)$ and the memory kernel
$\gamma (t)$ are given by

\begin{equation}\label{3}
\hat{F} (t) = \sum_j \left [ \left \{ \frac{\omega_j^2}{c_j}
\hat{x}_j (0) - \hat{q} (0) \right \} \frac{c_j^2}{\omega_j^2}
\cos \omega_j t + \frac{c_j}{\omega_j} \hat{p}_j (0) \sin
\omega_j t \right ]
\end{equation}

and

\begin{equation}\label{4}
\gamma (t) = \sum_{j=1}^N \frac{c_j^2}{\omega_j^2} \cos \omega_j
t \; \; ,
\end{equation}

Here masses have been assumed to be unity.

The Eq.(\ref{2}) is the well known exact quantized operator
Langevin equation for which the noise properties of $\hat{F} (t)$
can be derived by using a suitable initial canonical distribution
of the bath co-ordinate and momentum operators at $t=0$ as
follows;

\begin{eqnarray}
\langle \hat{F} (t) \rangle_{QS} &=& 0 \label{5} \\
\frac{1}{2} \{ \langle\hat{F}(t)\hat{F}(t^\prime)\rangle_{QS}&+&
\langle\hat{F}(t^\prime)\hat{F}(t)\rangle_{QS} \}\\ &=&
\frac{1}{2} \sum_{j=1}^N \frac{c_j^2}{\omega_j^2} \hbar \omega_j
\left(\coth \frac{\hbar\omega_j}{2 k_B T}\right) \cos
\omega_j(t-t^\prime)\label{6}
\end{eqnarray}

where $\langle...\rangle_{QS}$ refers to quantum statistical
average on bath degrees of freedom and is defined as

\begin{equation}\label{7}
\langle \hat{O} \rangle_{QS} = \frac{{\rm Tr} \hat{O} \exp
(-\hat{H}_{{\rm bath}}/k_BT)}{{\rm Tr}\exp (-\hat{H}_{{\rm
bath}}/k_BT)}
\end{equation}

for any operator $\hat{O}(\{(\omega_j^2/c_j)\hat{x}_j -
\hat{q}\},\{\hat{p}_j\})$ where $\hat{H}_{{\rm bath}}
(\sum^N_{i=1} (\hat{p}_i^2/2 + 1/2 (\omega_i \hat{x}_i -
\frac{c_i}{\omega_i} \hat{q} )^2))$ at $t=0$. By Trace we mean the
usual quantum statistical average. Eq.(\ref{6}) is the
fluctuation-dissipation relation with the noise operators ordered
appropriately in the quantum mechanical sense.

To construct a c-number Langevin equation we proceed from
Eq.(\ref{2}). We first carry out the {\it quantum mechanical
average} of Eq.(\ref{2})

\begin{equation}\label{8}
\langle \ddot{ \hat{q} } (t) \rangle + \int_0^t dt' \gamma (t-t')
\langle \dot{ \hat{q} } (t') \rangle + \langle V' ( \hat{q} )
\rangle = \langle \hat{F} (t) \rangle
\end{equation}

where the quantum mechanical average $\langle \ldots \rangle$ is
taken over the initial product separable quantum states of the
particle and the bath oscillators at $t=0$, $| \phi \rangle \{ |
\alpha_1 \rangle | \alpha_2 \rangle \ldots | \alpha_N \rangle \}
$. Here $| \phi \rangle$ denotes any arbitrary initial state of
the particle and $| \alpha_i \rangle$ corresponds to the initial
coherent state of the $i$-th bath oscillator. $|\alpha_i \rangle$
is given by $|\alpha_i \rangle = \exp(-|\alpha_i|^2/2)
\sum_{n_i=0}^\infty (\alpha_i^{n_i} /\sqrt{n_i !} ) | n_i \rangle
$, $\alpha_i$ being expressed in terms of the mean values of the
shifted co-ordinate and momentum of the $i$-th oscillator,
$\{(\omega_i^2/c_i)\langle \hat{x}_i (0) \rangle - \langle
\hat{q}(0) \rangle\} = ( \sqrt{\hbar} /2\omega_i) (\alpha_i +
\alpha_i^\star )$ and $\langle \hat{p}_i (0) \rangle = i
\sqrt{\hbar\omega_i/2 } (\alpha_i^\star - \alpha_i )$,
respectively. It is important to note that $\langle \hat{F} (t)
\rangle$ of Eq.(\ref{8}) is a classical-like noise term which, in
general, is a non-zero number because of the quantum mechanical
averaging and is given by $(\langle \hat{F}(t) \rangle \equiv
f(t))$;

\begin{equation}\label{9}
f(t) = \sum_j \left [ \left \{ \frac{\omega_j^2}{c_j} \langle
\hat{x}_j (0)\rangle - \langle\hat{q} (0)\rangle \right \}
\frac{c_j^2}{\omega_j^2} \cos \omega_j t + \frac{c_j}{\omega_j}
\langle\hat{p}_j (0)\rangle \sin \omega_j t \right ]
\end{equation}

It is convenient to rewrite the $c$-number equation (\ref{8}) as
follows;

\begin{equation}\label{10}
\langle \ddot{ \hat{q} } (t) \rangle + \int_0^t dt' \gamma (t-t')
\langle \dot{ \hat{q} } (t') \rangle + \langle V' ( \hat{q} )
\rangle = f (t)
\end{equation}

To realize $f(t)$ as an effective c-number noise we now introduce
the ansatz that the momenta $\langle \hat{p}_j (0) \rangle$ and
the shifted co-ordinates
$\{(\omega_j^2/c_j)\langle\hat{x}_j(0)\rangle -
\langle\hat{q}(0)\rangle\}$, $\{\hat{p}_j\}$ of the bath
oscillators are distributed according to a canonical distribution
of Gaussian form as

\begin{equation}\label{11}
{P}_j = {N} \exp \left\{ -\frac{[\langle \hat{p}_j(0) \rangle^2 +
\frac{c_j^2}{\omega_j^2} \{ \frac{\omega_j^2}{c_j}\langle
\hat{x}_j (0) \rangle - \langle \hat{q} (0) \rangle \}^2 ] }{ 2
\hbar \omega_j \left( \bar{n}_j(\omega_j) + \frac{1}{2} \right) }
\right\}
\end{equation}

so that for any function of the quantum mechanical mean values
$O_j\{\langle\hat{p}_j(0)\rangle,\\(({\omega_j^2}/{c_j})\langle\hat{x}_j
(0) \rangle  - \langle \hat{q} (0) \rangle )\}$ the statistical
average $\langle \ldots \rangle_S$ is

\begin{eqnarray}
\langle O_j \rangle_S & = &\int O_j \;{P}_j \; d\langle \hat{p}_j
(0) \rangle \;d \{ (\omega_j^2/c_j)\langle \hat{x}_j (0) \rangle -
\langle \hat{q} (0) \rangle \} \; \; . \label{12}
\end{eqnarray}

\noindent Here $\bar{n}_j$ indicates the average thermal photon
number of the $j$-th oscillator at temperature $T$ and
$\bar{n}_j(\omega_j) = 1/[\exp \left ( \hbar \omega_j/k_BT \right
) - 1]$ and ${N}$ is the normalization constant.

The distribution (\ref{11}) and the definition of statistical
average (\ref{12}) imply that $f(t)$ must satisfy

\begin{equation}\label{13}
\langle f(t) \rangle_S = 0
\end{equation}

and

\begin{equation}\label{14}
\langle f(t)f(t^{\prime})\rangle_S = \frac{1}{2} \sum_j
\frac{c_j^2}{\omega_j^2}\hbar \omega_j \left ( \coth \frac { \hbar
\omega_j }{ 2k_BT } \right ) \cos \omega_j (t - t^{\prime})
\end{equation}

\noindent That is, $c$-number noise $f(t)$ is such that it is
zero-centered and satisfies the standard fluctuation-dissipation
relation (FDR) as expressed in Eq.(\ref{6}). It is important to
emphasize that the ansatz (\ref{11}) is a canonical Wigner
distribution for a shifted harmonic oscillator \cite{hil} which
remains always a positive definite function. A special advantage
of using this distribution is that it remains valid as pure state
non-singular distribution function at $T = 0$. Furthermore, this
procedure allows us to {\it bypass the operator ordering}
prescription of Eq.(\ref{6}) for deriving the noise properties of
the bath in terms of fluctuation-dissipation relation and to
identify $f(t)$ as a classical looking noise with quantum
mechanical content.

We now return to Eq.(\ref{10}) to add the force term
$V^\prime(\langle\hat{q} \rangle)$ on both sides of Eq.(\ref{10})
and rearrange it to obtain

\begin{eqnarray}
\dot q &=& p
\label{15}\\
\dot p &=& - \int_0^t dt^\prime \gamma (t - t^\prime) p(t^\prime)
- V^\prime (q) + f(t) + Q(t) \label{16}
\end{eqnarray}

where we put $\langle\hat{q}(t)\rangle = q(t)$ and
$\langle\hat{p}(t)=p(t)$ for notational convenience and

\begin{equation}\label{17}
Q(t) = V^\prime(\langle\hat{q}\rangle) - \langle V^\prime(\hat{q})
\rangle
\end{equation}

represents the quantum correction due to the system degrees of
freedom. Eq.(\ref{16}) offers a simple interpretation. This
implies that the quantum Langevin equation is governed by a
$c$-number quantum noise $f(t)$ originating from the heat bath
characterized by the properties (\ref{13}) and (\ref{14}) and a
quantum fluctuation term $Q(t)$ \cite{db1,skb,db2,db3,bk1,bk2}
characteristic of the non-linearity of the potential \cite{tani}.

Referring to the quantum nature of the system in the Heisenberg
picture, one may writes.

\begin{eqnarray}
\hat{q} (t) &=& q + \delta\hat{q} \label{18}  \\
 \hat{p} (t) &=& p + \delta\hat{p}\label{19}
\end{eqnarray}

where $\langle\hat{q}\rangle(=q)$ and $\langle\hat{p}\rangle(=p)$
are the quantum-mechanical averages and $\delta\hat{q}$,
$\delta\hat{p}$ are the operators. By construction
$\langle\delta\hat{q}\rangle$ and $\langle\delta\hat{p}\rangle$
are zero and $[\delta\hat{q},\delta\hat{p}] = i\hbar$. Using
Eqs.(\ref{18}) and (\ref{19}) in $\langle V^{\prime} (\hat{q})
\rangle$ and a Taylor series expansion around
$\langle\hat{q}\rangle$ it is possible to express $Q(t)$ as

\begin{equation}\label{20}
Q(t) = -\sum_{n \ge 2} \frac{1}{n!} V^{(n+1)} (q)
\langle\delta\hat{q}^n(t)\rangle
\end{equation}

Here $V^{(n)}(q)$ is the n-th derivative of the potential $V(q)$.
For example, the second order $Q(t)$ is given by $Q(t) =
-\frac{1}{2} V^{\prime\prime\prime} (q) \langle \delta \hat{q}^2
\rangle$. The calculation of $Q(t)$
\cite{db1,skb,db2,db3,bk1,bk2,sm,akp} therefore rests on quantum
correction terms, $\langle \delta \hat{q}^n (t) \rangle$ which
are determined by solving the quantum correction equations as
discussed in the Sec.V.

The c-number Hamiltonian corresponding to Langevin equation
(\ref{15}, \ref{16}) is given by

\begin{eqnarray}
H = \frac{p^2}{2} &+& \left[ V(q) + \sum_{n \geq 2} \frac{1}{n!}
V^{(n)}(q) \langle {\delta \hat q^n} \rangle \right]\nonumber\\
& + & \sum^N_{i=1} \left \{ \frac{p_i^2}{2} + \frac{1}{2} (
\omega_i x_i - \frac{c_i}{\omega_i} q )^2 \right \}\label{21}
\end{eqnarray}

Note that the above Hamiltonian is different from our starting
Hamiltonian operator (\ref{1}) because of the c-number nature of
(\ref{21}). $\{x_i,p_i\}$ are the quantum mean value of the
co-ordinate and the momentum operators of the bath oscillators.

The spectral density function is defined as

\begin{equation}\label{24}
J(\omega)=\frac{\pi}{2} \sum_{i=1}^N \frac{c_i^2}{\omega_i}\;
\delta(\omega-\omega_i)
\end{equation}

We now assume that at $q=0$, the potential $V(q)$ has a barrier
with height $V^\ddag$ such that a harmonic approximation around
$q=0$ leads to

\begin{equation}\label{25}
V(q) = V^\ddag - \frac{1}{2} \omega_b^2 q^2 + V_2(q)
\end{equation}

where $\omega_b^2 = V''(q)\mid_{q=0}$, refers to the second
derivative of the potential $V(q)$. $\omega_b$ is the frequency at
the barrier top and $V_2(q)$ is the non-linear part of the
classical potential and is given by
$V_2=\sum_{n\ge3}\frac{1}{n!}\frac{\partial^nV(q)}{\partial
q^n}\mid_{q=0}q^n$. With Eq.(\ref{25}) the quantum correction part
in the Hamiltonian Eq.(\ref{21}) becomes

\begin{equation}\label{26}
\sum_{n\geq 2} \frac{1}{n!} V^{(n)}(q) \langle \overline{\delta
\hat q^n} \rangle = -\frac{\omega_b^2}{2} B_2 + V_3(q)
\end{equation}

where $B_n=\langle \overline{\delta \hat q^n} \rangle ;
V_3=\sum_{n \geq 2} \frac{B_n}{n!}
\frac{\partial^nV_2(q)}{\partial q^n}$. Note that we have
introduced an approximation by putting a bar over quantum
dispersion $\langle \delta \hat{q}^n (t) \rangle$ to indicate its
time average since we will be concerned here with the energy loss
of the system mode averaged over one round trip time,
\textit{i.e.}, the time required to traverse from one turning
point of the potential well to another and back. We will discuss
this averaging in greater detail in Sec.V. Putting (\ref{25}) and
(\ref{26}) in the Hamiltonian (\ref{21}) we obtain

\begin{equation}\label{27}
H = H_0 + V_N (q)
\end{equation}

where we have decomposed the Hamiltonian in the harmonic part
$H_0$ and the anharmonic part $V_N(q)$ as

\begin{equation}\label{28}
H_0 =\left[ \frac{p^2}{2} + \sum_i \frac{p_i^2}{2} \right] +
\left[ V_1^\ddag - \frac{1}{2} \omega_b^2 q^2 + \sum_i \frac{1}{2}
( \omega_i x_i - \frac{c_i}{\omega_i} q )^2 \right]
\end{equation}

and

\begin{eqnarray}
V_N(q) & = & V_2(q)+V_3(q)\label{29}\\
and\;\;\;\;\;
 V_1^\ddag & = & V^\ddag - \frac{B_2}{2} \omega_b^2\nonumber
\end{eqnarray}

$V_2(q)$ and $V_3(q)$ are therefore classical and quantum
anharmonic contributions to total anharmonic part of the
Hamiltonian. The separability of the c-number Hamiltonian in the
quadratic and nonlinear parts allows us to make a normal mode
transformation to convert the quadratic Hamiltonian into a
diagonal form. The method of normal mode analysis has been used
extensively by Pollak and co-workers in classical and quantum
theories of activation, tunneling and dephasing. For details we
refer to [Ref. 9,10,27-30].

Following Pollak, we diagonalize the force constant matrix $T$ of
the Hamiltonian (\ref{28}) with the matrix $U$

\begin{equation}\label{30}
U T = \lambda^2 U
\end{equation}

where $U$ provides the transformation from old co-ordinates to the
normal co-ordinates

\begin{eqnarray}
\left(
\begin{array}{cc}
{\rho}\\
{y_1}\\
{y_2}\\
{.}\\
{.}\\
{y_N}
\end{array}
\right) = U \left(
\begin{array}{cc}
{q}\\
{x_1}\\
{x_2}\\
{.}\\
{.}\\
{x_N}
\end{array}
\right)\label{31}
\end{eqnarray}

The c-number Hamiltonian of the unstable normal co-ordinate is
given by

\begin{equation}\label{32}
H_0 = \frac{1}{2} \dot \rho^2 + V_1^\ddag - \frac{1}{2}
\lambda_b^2 \rho^2 + \sum_{i=1}^N \frac{1}{2} \left( \dot y_i^2 +
\lambda_i^2 y_i^2 \right)
\end{equation}

The eigenvalues $\lambda_i^2$ and $\lambda_b^2$ are expressible
in terms of the coupling constant of the system and the bath
implicitly as follows:

\begin{eqnarray}
\lambda_b^2 & = & \omega_b^2 / \left[1 + \sum_{j=1}^N
\frac{c_j^2}{\omega_j^2 (\omega_j^2 + \lambda_b^2)} \right] \label{33}\\
\lambda_i^2 & = & - \omega_b^2 / \left[1 + \sum_{j=1}^N
\frac{c_j^2}{\omega_j^2 (\omega_j^2 - \lambda_i^2)} \right],\;\;\;
i=1,2...N \label{34}
\end{eqnarray}

where (\ref{33}) and (\ref{34}) correspond to normal mode
frequencies of the unstable mode and the i-th bath oscillator
respectively.

The transformation (\ref{31}) implies

\begin{equation}\label{35}
q = u_{00}\; \rho +\sum_{j=1}^N u_{j0} \;y_{j}
\end{equation}

and it has been shown \cite{pgh} that $u_{00}$ and $u_{j0}$ may be
expressed as

\begin{equation}\label{36}
u_{00}^2 = \left[ 1 + \sum_{j=1}^N \frac{c_j^2}{ (\omega_j^2 +
\lambda_b^2)^2} \right]^{-1}
\end{equation}

and

\begin{equation}\label{37}
u_{j0}^2 = \left[ 1 + \sum_{j=1}^N \frac{c_j^2}{ (\lambda_j^2 -
\omega_j^2)^2} \right]^{-1}
\end{equation}

Making use of the spectral density function (\ref{24}) and
Laplace transformation of $\gamma(t)$ Eq.(\ref{33}) and
Eq.(\ref{36}) may be written in the continuum limit as

\begin{equation}\label{38}
\lambda_b^2 = \frac{\omega_b^2}{1 + \widetilde{ \gamma}
(\lambda_b)/\lambda_b}
\end{equation}

and

\begin{equation}\label{39}
u_{00}^2 = \left[ 1 + \frac{2}{\pi} \int_0^\infty d\omega
\frac{J(\omega) \; \omega \; }{(\lambda_b^2 + \omega^2)^2}
\right]^{-1}
\end{equation}

The two important identities in relation to orthogonal
transformation matrices and the associated frequencies may be
noted here for the dynamics at the barrier top and at the bottom
of the well;

\begin{equation}\label{40}
\omega_b^2 \prod_{i=1}^N \omega_i^2 = \lambda_b^2 \prod_{i=1}^N
\lambda_i^2
\end{equation}

and

\begin{equation}\label{41}
\omega_0^2 \prod_{i=1}^N \omega_i^2 = \lambda_0^2 \prod_{i=1}^N
\Lambda_i^2
\end{equation}

Here $\omega_0$ and $\lambda_0$ are the frequencies of the system
at the bottom of the well in the original co-ordinate and normal
co-ordinate respectively. Similarly $\Lambda_i$ corresponds to the
normal mode frequencies of the bath oscillators coresponding to a
normal mode Hamiltonian at the bottom of the well,

\begin{equation}\label{x1}
H_0^\prime = \frac{1}{2} \dot \rho^{\prime 2} + \frac{1}{2}
\lambda_0^2 \rho^{\prime 2} + \left\{ \sum_{i=1}^N \frac{1}{2}
\dot y_i^{\prime 2} + \frac{1}{2} \Lambda_i^2\; y_i^{\prime 2}
\right\}
\end{equation}

Here $\rho^\prime$ and $\dot \rho^{\prime}$ are coordinate and
momentum of system mode respectively and $y_{i}^{\prime}$ and
$\dot y_{i}^{\prime}$ are coordinate and momentum of \textit{ith}
bath oscillator respectively at the bottom of the well in the
normal coordinates. $\lambda_0$ and $\Lambda_i$ are given by

\begin{eqnarray}
\lambda_0^2 & = & \omega_0^2 / \left[1 + \sum_{j=1}^N
\frac{c_j^2}{\omega_j^2 (\omega_j^2 - \lambda_0^2)} \right] \label{x2}\\
\Lambda_i^2 & = & \omega_0^2 / \left[1 + \sum_{j=1}^N
\frac{c_j^2}{\omega_j^2 (\omega_j^2 - \Lambda_j^2)} \right],\;\;\;
i=1,2...N \label{x13}
\end{eqnarray}

The description of a Kramers' turnover of the rate constant from
low to high friction limit requires the coupling between the
normal modes. An important parameter which is relevant for
describing the coupling perturbatively has been defined by

\begin{equation}\label{42}
\epsilon = \left(\frac{1}{u_{00}^2} - 1 \right) = \sum_{j=1}^N
\left(\frac{u_{j0}}{u_{00}}\right)^2
\end{equation}

which can be expressed further in terms of the frequency dependent
friction as

\begin{equation}\label{43}
\epsilon = \frac{1}{2} \left[
\frac{\widetilde{\gamma}(\lambda_b)}{\lambda_b} + \frac{\partial
\widetilde{\gamma}(\lambda_b)}{\partial \lambda_b} \right]
\end{equation}

Furthermore for future use we also define a parameter

\begin{equation}\label{44}
K_c(t) = \sum_{i=1}^N \left( \frac{u_{i0}}{u_{00}} \right)^2 \cos
(\lambda_i t)
\end{equation}

whose Laplace transform is given by \cite{pgh,p3}

\begin{equation}\label{45}
\widetilde{K}_c(s) = \frac{1}{u_{00}^2}\; \frac{s}{s^2+s\;
\widetilde{\gamma}(s)-\omega_b^2}-\frac{s}{s^2-\lambda_b^2}
\end{equation}

\section{{\bf C-number quantum version of multidimensional TST}}

To begin with we consider the particle to be trapped in a well
described by a potential $V(q)$ depicted schematically in Fig.1.
In the normal mode description of $(N+1)$ oscillators according
to the Hamiltonian (\ref{32}) the bath modes and the system mode
are uncoupled. Considering the unstable reaction co-ordinate to be
thermalized according to the Wigner thermal canonical
distribution of $N$ uncoupled harmonic oscillators plus one
inverted we have

\begin{equation}\label{46}
W_{eq} = z^{-1} \exp \left[ - \;\frac{\frac{1}{2} \dot \rho^2 +
V_1^\ddag - \frac{1}{2} \lambda_b^2 \rho^2}{\hbar \lambda_0
(\overline{n}_0(\lambda_0) + \frac{1}{2})} \right] \prod_{i=1}^N
\exp \left[ - \;\frac{\frac{1}{2} \dot y_i^2 + \frac{1}{2}
\lambda_i^2 y_i^2}{\hbar \lambda_i (\overline{n}_i(\lambda_i) +
\frac{1}{2})} \right]
\end{equation}

$z$ is the normalization constant. As usual this can be
calculated using the distribution function inside the reactant
well. For this it is necessary to consider the normal mode
Hamiltonian at the bottom of the well expressed as $H_0^{\prime}$
in the Eq.(\ref{x1}). The corresponding distribution in the well
is

\begin{equation}\label{48}
W_{eq} = z^{-1} \exp \left[ -\; \frac{\frac{1}{2} \dot
\rho^{\prime 2} + \frac{1}{2} \lambda_0^2 \rho^{\prime 2}}{\hbar
\lambda_0 (\overline{n}_0(\lambda_0) + \frac{1}{2})} \right]
\prod_{i=1}^N \exp \left[  -\; \frac{\frac{1}{2} \dot y_i^{\prime
2} + \frac{1}{2} \Lambda_i^2 y_i{\prime ^2}}{\hbar \Lambda_i
(\overline{n}_i(\Lambda_i) + \frac{1}{2})} \right]
\end{equation}

which can be normalized to obtain

\begin{equation}\label{49}
z^{-1} = \frac{\lambda_0}{ 2 \pi \hbar \lambda_0
(\overline{n}_0(\lambda_0) + \frac{1}{2})} \; \prod_{i=1}^N
\frac{\Lambda_i}{ 2 \pi \hbar \Lambda_i (\overline{n}_i(\Lambda_i)
+ \frac{1}{2})}
\end{equation}

The identity relation (\ref{41}) can be used to transform
(\ref{49}) to the following form

\begin{equation}\label{50}
z^{-1} = \frac{\omega_0}{ 2 \pi \hbar \lambda_0
(\overline{n}_0(\lambda_0) + \frac{1}{2})} \; \prod_{i=1}^N
\frac{\omega_i}{ 2 \pi \hbar \Lambda_i (\overline{n}_i(\Lambda_i)
+ \frac{1}{2})}
\end{equation}

Putting Eq.(\ref{50}) in Eq.(\ref{46}) we obtain after
integration over the stable modes

\begin{equation}\label{51}
W_{eq} = \frac{\omega_0}{ 2 \pi \hbar \lambda_0
(\overline{n}_0(\lambda_0) + \frac{1}{2})}\;
\frac{\lambda_b}{\omega_b}\;\chi\; \exp \left[ - \frac{\frac{1}{2}
\dot \rho^2 + V_1^\ddag - \frac{1}{2} \lambda_b^2 \rho^2}{\hbar
\lambda_0 (\overline{n}_0(\lambda_0) + \frac{1}{2})} \right]
\end{equation}

where

\begin{equation}\label{x3}
\chi = \prod_{i=1}^{N}\frac{\hbar \lambda_i
(\overline{n}_i(\lambda_i) + \frac{1}{2})}{\hbar \Lambda_i
(\overline{n}_i(\Lambda_i) + \frac{1}{2})}
\end{equation}

The total energy of the unstable mode is

\begin{equation}\label{52}
E = \frac{1}{2} \dot \rho^2 + V_1^\ddag - \frac{1}{2} \lambda_b^2
\rho^2
\end{equation}

The prime quantity for determination of rate constant is the
distribution of energy of the unstable mode. Thus going over to
an energy space so that the co-ordinate $\rho, \;\dot \rho$ are
transformed to $t, E$, respectively with unit Jacobian, the
equilibrium distribution function (\ref{51}) is given by,

\begin{equation}\label{53}
f_{eq}(E) = \frac{\omega_0}{ 2 \pi \hbar \lambda_0
(\overline{n}_0(\lambda_0) + \frac{1}{2})}\;
\frac{\lambda_b}{\omega_b}\;\chi\; \exp \left[ - \frac{E}{\hbar
\lambda_0 (\overline{n}_0(\lambda_0) + \frac{1}{2})} \right]
\end{equation}

The above distribution is valid for the energy of the unstable
mode $E>V^\ddag$ as well as $E<V^\ddag$.

\subsection{\underline{Quantum multidimensional TST rate}}

The rate of activated barrier crossing in terms of the equilibrium
probability becomes

\begin{equation}\label{54}
\Gamma = \int_{V^\ddag}^\infty f(E)\; dE
\end{equation}

As the unstable mode remains uncoupled from the stable modes the
former mode behaves deterministically and the recrossing does not
occur in this case.

Making use of the distribution (\ref{53}) in (\ref{54}) we obtain
the rate constant

\begin{equation}\label{55}
\Gamma_{TST} = \frac{\omega_0}{2
\pi}\;\frac{\lambda_b}{\omega_b}\;\chi\;\exp \left[ -\;
\frac{V^\ddag}{\hbar \lambda_0 (\overline{n}_0(\lambda_0) +
\frac{1}{2})} \right]
\end{equation}

The above expression corresponds to a quantum multidimensional
transition state rate constant. This is central result of this
section. Apart from usual Kramers'-Grote-Hynes term
$\lambda_b/\omega_b$ and $\omega_0/2\pi$, the term arising out of
classical transition state result, Eq.(3.10) contains two
important factors. First, an exponential Arrhenius term where the
usual thermal factor $k_BT$ is replaced by $\hbar \lambda_0
(\overline{n}_0(\lambda_0) + \frac{1}{2})$ includes quantum
effects due to heat bath at very low temperature. In the high
temperature limit it reduces to $k_BT$ and one recovers the usual
Boltzmann factor. This term is essentially an offshoot of a
description of thermal equilibrium by a canonical Wigner
distribution of harmonic oscillators heat bath. Second term
$\chi$ can be identified as the quantum correction to Grote-Hynes
factor or more precisely a vacuum corrected generalized Wolynes
contribution for quantum transmission and reflection for the
finite barrier. While usual Wolynes term takes care of the
quantum effects at the higher temperature the factor $\chi$
incorporates quantum effects at arbitrary low temperature. In
what follows we show that the usual Wolynes term can easily be
recovered for $\chi$ in the appropriate limit.

\subsection{\underline{Derivation of Wolynes factor from $\chi$}}

We begin by noting that $\overline{n}(x)$ in $\chi$ which is
given by

\begin{equation}\label{x4}
\chi = \prod_{i=1}^{N}\frac{\hbar \lambda_i
(\overline{n}_i(\lambda_i) + \frac{1}{2})}{\hbar \Lambda_i
(\overline{n}_i(\Lambda_i) + \frac{1}{2})}
\end{equation}

is the Bose distribution $\overline{n}(x)=(e^{\hbar
x/k_BT}-1)^{-1}$. Neglecting the vacuum contribution $1/2$ from
the terms like $\hbar x(\overline{n}(x)+\frac{1}{2})$ and keeping
only the leading order quantum contribution in the thermal limit
we obtain

\begin{equation}\label{x5}
\hbar x\left(\overline{n}(x)+\frac{1}{2}\right) \approx
\frac{\hbar x}{2} \left(\sinh \frac{\hbar x}{2 k_BT}\right)^{-1}
\end{equation}

Therefore $\chi$ reduces to $\Xi$ (say) the Wolynes factor

\begin{equation}\label{x6}
\Xi = \prod_{i=1}^{N}\frac{\lambda_i \left(\sinh \frac{\hbar
\lambda_i }{2k_BT}\right)^{-1}}{\Lambda_i \left(\sinh \frac{\hbar
\Lambda_i }{2k_BT}\right)^{-1}}
\end{equation}

From identities (\ref{40}) and (\ref{41}) it follows

\begin{eqnarray}
\prod_{i=1}^{N}\lambda_{i}=\frac{\omega_b}{\lambda_b}\prod_{i=1}^{N}\omega_i\label{x7}\\
\prod_{i=1}^{N}\Lambda_{i}=\frac{\omega_0}{\lambda_0}\prod_{i=1}^{N}\omega_i\label{x8}
\end{eqnarray}

respectively, and we have
\begin{equation}
\prod_{i=1}^{N}\frac{\lambda_i}{\Lambda_{i}}=\frac{\lambda_0
\omega_b}{\omega_0 \lambda_b}\label{x15}
\end{equation}

Making use of the relation (\ref{x15}) in (\ref{x6}) we obtain

\begin{equation}\label{x9}
\Xi = \frac{\lambda_0 \omega_b}{\omega_0 \lambda_b}
\prod_{i=1}^{N}\frac{\sinh ({\hbar \Lambda_i }/{2k_BT})}{\sinh
({\hbar \lambda_i }/{2k_BT})}
\end{equation}

Furthermore $({\lambda_0 \omega_b})/({\omega_0 \lambda_b})$ can
be rewritten as $\frac{\omega_b}{\omega_0}\frac{(\hbar
\lambda_0/k_BT)}{(\hbar \lambda_b/k_BT)}$ which may be
approximated in the form $(\omega_b/\omega_0)\frac{\sinh(\hbar
\lambda_0/2k_BT)}{\sinh(\hbar \lambda_b/2k_BT)}$. Eq.(3.17) then
reduces to

\begin{equation}\label{x10}
\Xi = \frac{\omega_b}{\omega_0}\;\frac{\sinh(\hbar
\lambda_0/2k_BT)}{\sinh(\hbar
\lambda_b/2k_BT)}\;\prod_{i=1}^{N}\frac{\sinh ({\hbar \Lambda_i
}/{2k_BT})}{\sinh ({\hbar \lambda_i }/{2k_BT})}
\end{equation}

This is the wellknown Wolynes \cite{rip,wolyn} expression derived
in early eighties as a higher temperature equilibrium quantum
correction to Kramers'-Grote-Hynes dynamical factor to Kramers'
rate. Both $\chi$ and the Wolynes factor become unity in the
classical limit. We conclude by noting that unlike Wolynes factor
$\Xi$, $\chi$ is valid below cross-over temperature.

\section{{\bf The coupling between the normal modes, energy loss and the quantum rate}}

The rate formula (\ref{54}) is quite general and requires the
knowledge of $f(E)$. In the last section the quantum version of
Kramers'-Grote-Hynes estimation of this rate rests on the
replacement of $f(E)$ by an equilibrium Wigner distribution
function in phase space. A determination of $f(E)$ from the
dynamics of the energy diffusion process, however is based on the
solution of the integral equation originally formulated by
Melnikov \cite{mel} and subsequently by others \cite{pgh,co},

\begin{equation}\label{56}
f(E) = \int_0^{V^{\ddag}} dE^{\prime}\; P(E|E^{\prime})\;
f(E^{\prime})
\end{equation}

which implies that $f(E)$ can be determined by a conditional
probability function $P(E|E^{\prime})\;dE$  that a system
escaping the barrier region with energy $E^\prime$ of the unstable
mode makes a round trip of the barrier with an energy between $E$
and $E+dE$. Deep inside the well, the strong coupling between the
stable and unstable modes brings $f(E)$ close to its equilibrium
value $W_{eq}$, whereas a weak coupling prevails where the energy
is close to the barrier energy. A detailed consideration of this
weak coupling between the stable and unstable modes results in
the conditional probability function $P(E|E^\prime)$ which when
made use in (\ref{56}) gives the rate constant beyond the
multidimensional TST limit.

To proceed further we construct the equation of motion for the
normal modes corresponding to the Hamiltonian (\ref{32}) using
(\ref{27}) and (\ref{35}).

\begin{eqnarray}
\ddot{\rho} - \lambda_b^2 \;\rho & = & - u_{00}\; V_N^{\prime}
(u_{00} \rho
+ \sum_{i=1}^N u_{i0}\; y_i)\label{57}\\
\ddot{y_i} + \lambda_i^2 \;y_i & = & - u_{i0}\; V_N^{\prime}
(u_{00} \rho + \sum_{i=1}^N u_{i0} \;y_i)\label{58}
\end{eqnarray}

where $V_N^\prime$ denotes the derivative of $V_N(x)$ with respect
to $x$. An important point is noteworthy. In contrast to
classical theories where $V_N$ is purely an anharmonic classical
contribution, $V_2$, the present treatment incorporates quantum
effects due to nonlinearity of the system potential $V_3$
entangled with quantum dispersion terms as given in
Eq.(\ref{29}). Since the coupling between the stable and unstable
normal modes is very sensitive to the small variation of energy
we expect the quantum effect to contribute significantly to the
energy loss mechanism and hence the calculation of depopulation
factor. This is an important point of departure from the earlier
treatment of Rips and Pollak who had considered the unstable mode
to be classical and neglected this contribution. The
applicability of the theory is thus not restricted to socalled
activated tunneling region  only.

Defining a smallness parameter $g_i$ as $g_i=u_{i0}/u_{00}$ and
$\epsilon=\sum g_i^2$ we may write the zero order equation of
motion for the unstable mode as

\begin{equation}\label{59}
\ddot{\rho} - \lambda_b^2 \;\rho = - u_{00}\; V_N^{\prime}
(u_{00}\;\rho)
\end{equation}

and for the stable modes as

\begin{equation}\label{60}
\ddot{y_i} + \lambda_i^2 \;y_i = g_i\; \zeta
\end{equation}

where $\zeta$ is a forcing function given by

\begin{equation}\label{61}
\zeta(t) = - u_{00}\; V_N^\prime (u_{00}\;\rho)
\end{equation}

The parameter $\epsilon\ll 1$ is for weak coupling. The forcing
function $\zeta(t)$ can be determined by the unperturbed motion
of the unstable mode $\rho$. The general solution of
Eq.(\ref{60}) is given by

\begin{equation}\label{62}
y_i(t) = y_i(0) \cos \lambda_i t + \frac{\dot y_i(0)}{\lambda_i}\;
\sin \lambda_i t + \int_0^t dt^\prime \;\frac{\sin \lambda_i
(t-t^\prime)}{\lambda_i}\; g_i\; \zeta(t^\prime)
\end{equation}

As noted earlier that at equilibrium the initial distribution of
the stable normal modes $y_i(0)$ and $\dot y_i(0)$ is a thermal
canonical Wigner distribution so that

\begin{eqnarray}
\langle y_i(0) \rangle_{S} = \langle \dot y_i(0) \rangle_{S} =
\langle y_i(0) \dot y_i(0) \rangle_{S}=0\nonumber\\
\langle \dot y_i(0)^2 \rangle_{S} = \lambda_i^2 \langle y_i(0)^2
\rangle_{S} = \hbar \lambda_i\; \left(\overline{n}_i(\lambda_i) +
\frac{1}{2}\right)\label{63}
\end{eqnarray}

The total c-number energy of the stable bath modes during the
traversal of $\rho$ mode over a round trip time $t_p$,

\begin{equation}\label{64}
E\; = \;\sum_{i=1}^N E_i\; = \;\sum_i \frac{1}{2}\; \dot y_i^2 +
\frac{1}{2}\; \lambda_i^2\; y_i^2
\end{equation}

can be calculated form (\ref{62}) and may be put in the form

\begin{equation}\label{65}
E\;=\;E^\prime - \Delta E + \delta E
\end{equation}

$E^\prime=\sum_i E_i(0)$ where $E_i(0)$ is the initial energy of
the i-th  bath oscillator. The gain of energy by the stable modes
is equal to the loss of energy of the unstable mode. This is
characterized by a systematic and a stochastic contribution to
energy loss of the unstable mode $\Delta E$ and $\delta E$,
respectively. These are given by

\begin{equation}\label{66}
\Delta E = \frac{1}{2}\; \sum_{i=1}^N g_i^2 \int_0^{t_p} dt
\int_0^{t_p} dt^\prime\; \cos [\lambda_i (t-t^\prime)]\;
\zeta(t)\; \zeta(t^\prime)
\end{equation}

\begin{equation}\label{67}
\delta E = \sum_{i=1}^N g_i \int_0^{t_p} dt\; [ \;\dot y_i(0)\;
\cos(\lambda_i t) - y_i(0) \;\lambda_i\; \sin(\lambda_i t)\;]
\;\zeta(t)
\end{equation}

While $\Delta E$ represents a systematic energy loss due to
coupling, $\delta E$ is a measure of instantaneous loss induced
by Gaussian fluctuation around $\langle E \rangle = E^\prime -
\Delta E$ such that

\begin{equation}\label{68}
\langle \delta E \rangle = 0 \;\;\;and\;\;\; \langle \delta E^2
\rangle = D
\end{equation}

Making use of the relation (\ref{63}), $\langle \delta E^2
\rangle$ can be calculated as

\begin{equation}\label{69}
\langle \delta E^2 \rangle = \sum_{i=1}^N g_i^2 \int_0^{t_p} dt
\int_0^{t_p} dt^\prime\; \hbar \lambda_i\;
\left[\overline{n}_i(\lambda_i) + \frac{1}{2}\right]\; \cos
[\lambda_i (t-t^\prime)]\; \zeta(t)\; \zeta(t^\prime)
\end{equation}

In the Markovian limit (\ref{69}) reduces to

\begin{equation}\label{70}
\langle \delta E^2 \rangle =2\; \hbar \lambda_b\;
\left[\overline{n}_b + \frac{1}{2}\right]\; \Delta E
\end{equation}

The mean and the second moments of the energy fluctuations lead
us to the conditional distribution function of the Gaussian form
as

\begin{equation}\label{71}
P(E|E^\prime) = \frac{1}{\sqrt{2 \pi D}}\; \exp \left[ -\;
\frac{(E-E^\prime+\Delta E)^2}{2 D} \right]
\end{equation}

The Markovian limit of the above expression can be obtained by
replacing $D$ by (\ref{70}).

We now proceed to calculate the rate explicitly. As is wellknown
that as probability function $f(E)$ tends to its equilibrium
limit $W_{eq}$ the rate $\Gamma$ reduces to the rate calculated
by multidimensional transition state theory. If, in general, they
differ by a depopulation factor $f_T$ as defined in

\begin{equation}\label{72}
\Gamma=f_T\;\frac{\omega_0}{2
\pi}\;\chi\;\frac{\lambda_b}{\omega_b}\;\exp \left[ -\;
\frac{V^\ddag}{\hbar \lambda_0 (\overline{n}_0(\lambda_0) +
\frac{1}{2})} \right]
\end{equation}

so that by comparing (\ref{72}) with (\ref{54}) to include all
energies we write

\begin{equation}\label{73}
f_T=\frac{2
\pi}{\omega_0}\;\frac{1}{\chi}\;\frac{\omega_b}{\lambda_b}\;\exp
\left[ \frac{V^\ddag}{\hbar \lambda_0 (\overline{n}_0(\lambda_0) +
\frac{1}{2})} \right] \int_0^\infty f(E)\; dE
\end{equation}

Introducing one-sided Fourier transformation ansatz as

\begin{eqnarray}
\phi^+(\lambda)= & \frac{2\pi}{\omega_0} &
\frac{1}{\chi}\;\frac{\omega_b}{\lambda_b}\;\exp \left[
\frac{V^\ddag}{\hbar \lambda_0 (\overline{n}_0(\lambda_0) +
\frac{1}{2})} \right]\nonumber\\
&\times& \int_0^\infty f(E) \exp\left[ \frac{2 (i \lambda + 1/2 )
\Delta E .E }{D}\right] \; dE \label{74}
\end{eqnarray}

and

\begin{eqnarray}
\phi^-(\lambda)= & \frac{2\pi}{\omega_0} &
\frac{1}{\chi}\;\frac{\omega_b}{\lambda_b}\;\exp \left[
\frac{V^\ddag}{\hbar \lambda_0 (\overline{n}_0(\lambda_0) +
\frac{1}{2})} \right]\nonumber\\
&\times& \int_{-\infty}^0 f(E) \exp\left[ \frac{2 (i \lambda + 1/2
) \Delta E .E }{D}\right] \; dE\label{75}
\end{eqnarray}

it is possible to transform (\ref{56}) into Wiener-Hopf
\cite{mel,co,ti,sv} equation with symmetric kernel which can be
solved by standard technique first suggested by Melnikov and
Meshkov in the rate theoretical context and subsequently used by
others. The ultimate expression of rate can be worked out as

\begin{eqnarray}
\Gamma=&\frac{\omega_0}{2\pi}&\frac{\lambda_b}{\omega_b}\;{\chi}\;\exp
\left[ -\; \frac{V^\ddag}{\hbar \lambda_0
(\overline{n}_0(\lambda_0) + \frac{1}{2})} \right]\nonumber\\
&\times&\exp\left[ \frac{1}{2 \pi} \int_{-\infty}^\infty
\frac{\ln[1-\exp(-\;\frac{2 \Delta E^2\; (\lambda^{\prime 2}+1/4)
}{D})]}{[\lambda^{\prime 2}+1/4]}\;d\lambda^\prime\right]\;
\label{76}
\end{eqnarray}

This is the central result of the paper. The quantum rate is a
product of several terms. The first term corresponds to classical
well frequency factor according to transition state theory. The
second is the classical Kramers'-Grote-Hynes factor for arbitrary
frequency dependent friction. The third term $\chi$ as given by
(\ref{x3}) corresponds to equilibrium quantum correction to
Kramers'-Grote-Hynes factor which reduces to high temperature
quantum correction or Wolynes term in the appropriate limit as
shown in Sec.III. The fourth term refers to the Wigner's canonical
thermal distribution in the harmonic well and reduces to the usual
Arrhenius factor in the classical limit when $\hbar \lambda_0
(\overline{n}_0(\lambda_0) + \frac{1}{2})\rightarrow k_{B}T$ for
$\hbar \lambda_0\ll k_{B}T$ This term takes care of the quantum
effects of the heat bath which thermalizes the particle inside
the well. The vacuum term $1/2$ prevents the distribution from
being singular as one approaches to absolute zero. Therefore a
significant contribution of quantum correction due to heat bath to
the rate enters through both equilibrium Wigner function and
$\chi$. The last term, the quantum depopulation factor is
essentially due to the weak coupling of the c-number unstable
normal mode with the stable modes at the barrier top. This
quantity signifies the nonequilibrium nature of the dynamics at
the barrier top and depends on the average energy loss of the
c-number unstable mode as well as on the energy dispersion. Since
both of them are sensitive to the coupling , we emphasize that
the quantum effect due to nonlinearity of the system makes its
presence felt in these quantities. Apart from these couplings,
quantum nature of the system is also manifested in $t_p$, the
round trip time, for a complete traversal in the well. The latter
point can be understood more explicitly as we go over to Section
VI.

Before closing this section we point out that (i) because of low
temperature quantum correction to Kramers'-Grote-Hynes factor, the
quantum nature of the unstable mode, and a Wigner description of
probability distribution inside the well, the rate expression is
valid both above and below the cross-over temperature. Thus this
works well below the socalled activated tunneling regime down to
vacuum limit. (ii) Since in this present calculation the
dissipation effects are not restricted to Markovian limit, the
rate expression is valid for arbitrary frequency-dependent
friction. (iii) Finally from the expressions for the two primary
quantities that determine the quantum depopulation factor ,
\textit{i.e.} , the average energy loss $\Delta E$ and energy
dispersion $\langle \delta E^2 \rangle$ it is apparent that the
classical limit of the depopulation factor can be recovered by
reducing these two quantities in the limit $k_BT\gg \hbar
\lambda_0$ and neglecting quantum correction in $V_N$. The result
is the classical depopulation factor \cite{pgh}. (iv) An important
advantage of the present c-number scheme is that it allows us to
use a Gaussian kernel (\ref{71}) which is exactly
\textit{classical in form} (as used by Melnikov and Meshkov)
\textit{but quantum mechanical in its content} since the energy
loss $\Delta E$ and the dispersion $[\langle \delta E^2
\rangle\equiv D]$ are quantum mechanical in character. The
quantum-classical correspondence thus be immediately restored.
The rate expression (\ref{76}) reduces to its classical
non-Markovian expression at high temperature where $\hbar x
(\overline{n}(x)+1/2)$ becomes $k_BT$ so that the Wolynes factor
$\chi$ goes over to unity and the depopulation factor takes its
classical value. We obtain the expression for
Pollak-Grabert-H\"{a}nggi as follows:

\begin{eqnarray}
\Gamma_{classical}=&\frac{\omega_0}{2\pi}&\frac{\lambda_b}{\omega_b}\;\exp
\left( -\; \frac{V^\ddag}{k_BT} \right)\nonumber\\
&\times&\exp\left[\frac{1}{2 \pi} \int_{-\infty}^\infty
\frac{\ln[1-\exp(-\;\frac{\Delta E}{k_BT}\; (\lambda^{\prime
2}+1/4))]}{[\lambda^{\prime 2}+1/4]}\;d\lambda^\prime\right]
\label{x12}
\end{eqnarray}

where $\Delta E$ as given by (\ref{66}) quantum corrections
arising out of system nonlinearity.

\section{{\bf  The quantum correction to energy loss of the unstable mode}}

The key issue in Kramers' turnover problem is the calculation of
energy loss $\Delta E$ of the unstable mode during its round trip
in the well over a time $t_p$. This is extremely sensitive to
coupling of the stable and the unstable modes described in terms
of the equations of motion for the normal modes (\ref{59}) and
(\ref{60}). The present calculation by virtue of considering the
quantum system mode takes care of this loss through $\zeta(t)$ or
through the nonlinearity of the system potential $V_N$.
Furthermore the time $t_p$ depends explicitly on the quantum
corrections as may be seen in the next section. Thus the quantum
corrections $B_n$ (in Eq.(\ref{26})) are the key quantities that
need to be determined. In an illustrative example of the next
section with a cubic potential, we show that $B_2$ is the relevant
quantity for the leading order quantum correction. In the
following we give a recipe for calculation of quantum
corrections. We return to the operator equation (\ref{2}) and put
(\ref{18}) and (\ref{19}) to obtain

\begin{equation}\label{77}
\delta \dot {\hat q}=\delta \hat p
\end{equation}

\begin{eqnarray}
\delta \dot {\hat p} &+& \int_0^t \gamma(t-t^\prime) \;\delta \hat
p(t^\prime)\; dt^\prime + V^{\prime \prime} (q)\; \delta \hat
q\nonumber\\ &+& \sum_{n\geq 2}\frac{1}{2} \;V^{(n+1)}(q)\;(
\delta \hat q^n-\langle\delta \hat q^n\rangle)
=\hat{F}(t)-f(t)\label{78}
\end{eqnarray}

We then perform a quantum mechanical averaging over bath states
with $\prod_{i=1}^N \{ | \alpha_{i}(0) \rangle \}$  to get rid of
the term $\hat F(t)-f(t)$. The Eqs.(\ref{77}) and (\ref{78}) along
with (\ref{15}) and (\ref{16}) form the key element for
calculation of the quantum mechanical correction due to the
nonlinearity of the potential. Considering the friction kernel
$\gamma(t)$ to be arbitrary (but decaying) we may calculate the
leading order quantum correction for the harmonic mode at the
barrier top for which higher derivatives of $V(q)$ in (\ref{78})
vanish. The above equations can then be solved by Laplace
transformation technique to obtain

\begin{equation}\label{79}
\delta {\hat q}(t)=\delta \hat p(0)\; C_v(t) + \left( 1 +
\omega_b^2 \int_0^t \;C_v(t^\prime) dt^\prime \right) \delta {\hat
q}(0)
\end{equation}

and

\begin{equation}\label{80}
\delta {\hat p}(t)=\delta \hat p(0)\; C_v(t) + \delta {\hat
q}(0)\; C_q(t)
\end{equation}

where

\begin{equation}\label{81}
C_v(t)=L^{-1} \left[ \frac{1}{s^2+s\; \widetilde{\gamma}(s) -
\omega_b^2} \right]
\end{equation}

and

\begin{equation}\label{82}
C_q(t) = 1 + \omega_b^2 \int_0^t C_v(t^\prime) \;dt^\prime
\end{equation}

and $\widetilde{\gamma}(s)$ is the Laplace transform of
$\gamma(t)$ defined as $\widetilde{\gamma}(s)=\int_0^\infty
\gamma(t) e^{-s t} dt$. After squaring and quantum mechanical
averaging Eq.(\ref{79}) yields

\begin{eqnarray}
\langle \delta \hat q^2(t) \rangle &=& \langle \delta \hat
p^2(0)\; \rangle C_v^2(t) + \langle \delta \hat q^2(0)\; \rangle
C_q^2(t)\nonumber\\ &+& C_v(t)\; C_q(t) \langle \delta \hat p(0)\;
\delta \hat q(0)+\delta \hat q(0)\; \delta \hat p(0)
\rangle\label{83}
\end{eqnarray}

The relevant quantum correction in $V_N$ in the leading order is
$B_2$ obtained as a time average of $\langle \delta \hat q^2(t)
\rangle$.

\begin{equation}\label{84}
B_2 = \frac{1}{t_p} \int_0^{t_p} \langle \delta \hat q^2(t)
\rangle \;dt
 \end{equation}

While for Markovian friction, the above equation can be
calculated analytically , one must take resort to numerical
evaluation of $C_v(t)$ and $C_q(t)$ for arbitrary friction kernel.
We emphasize that the correction $B_2$ is the leading order
quantum correction for the unstable mode and the friction kernel
is considered to be arbitrary in nature. Thus the quantum
correction due to the unstable mode affects $\langle\delta
E^2\rangle$ as well as $\Delta E$ through $t_p$ and $\zeta(t)$.
The quantum nature of the heat bath on the other hand is taken
care of through the width parameter of the Wigner canonical
thermal distribution function. It is pertinent to mention at this
point that the quantum correction to the average energy loss and
dispersion of energy can be calculated by including higher order
contribution with the help of the basic equations (\ref{77}) and
(\ref{78}) within the framework of the theory. However, the
coupling of the unstable and the stable modes at the barrier top
being weak it is sufficient to consider this leading order
quantum correction for the present treatment.

\section{{\bf Application to a cubic oscillator}}

The theory developed so far is fairly general in the sense that
it takes into account of an arbitrary form of nonlinear potential
$V(q)$ and a frequency dependent friction. We now consider a
simple nonlinear potential of the form chosen for comparison with
standard result \cite{pgh,rip}

\begin{equation}\label{85}
V(q) = - a q^3 -b q^2 + \frac{4 b^3}{27 a^2}
\end{equation}

The extrema of the potential corresponds to $q=0$ and
$q=(-q_0)=-2 b/3 a$, the respective potentials being $4 b^3/27
a^2$ and $0$ respectively. Thus we have

\begin{equation}\label{86}
V^\ddag = \frac{4 b^3}{27 a^2},\;\;\;\omega_0^2=2
b,\;\;\;\omega_b^2=2 b
\end{equation}

the metastable minima is at $q=-q_0$. Furthermore we assume a
Lorentian form of spectral density function, $J(\omega)$ as

\begin{equation}\label{87}
J(\omega) = \frac{\omega\; \gamma}{1 + \omega^2\; \gamma^2\;
\tau_c^2}
\end{equation}

for which the friction kernel has an exponential form as

\begin{equation}\label{88}
\gamma (t) = \frac{1}{\tau_c}\; \exp ( - \frac{t}{\gamma \tau_c})
\end{equation}

The Laplace transform of $\gamma (t)$ is given by

\begin{equation}\label{89}
\widetilde{\gamma} (s) = \gamma /(1 + s \;\gamma \;\tau_c )
\end{equation}

We follow the classical procedure of Straub, Borkovec and Berne
\cite{s1,bj} to consider the potential (\ref{85}) in the form of a
piecewise continuous harmonic potential as

\begin{eqnarray}
V(q) & \approx & \frac{1}{2}\; \omega_0^2\;( q + q_0)^2\;\;\;\;
for\;\;\;q<-q^*\label{90}\\
\;\;\;\; & \approx & V_a^\ddag - \frac{1}{2}\;
\omega_b^2\;q^2\;\;\;\;for\;\;\;q\ge-q^*\label{91}
\end{eqnarray}

From the continuity of the potential $V(q)$ and its derivatives
at $q=-q^*$, \textit{i.e.}, $1/2\; \omega_0^2\;( q +
q_0)^2|_{-q^*} = V_a^\ddag - 1/2\; \omega_b^2\;q^2|_{-q^*}$ and
$\omega_0^2\;( q + q_0)|_{-q^*} = - \omega_b^2\;q|_{-q^*}$ we
obtain

\begin{equation}\label{92}
V_a^\ddag = \frac{2 b^3}{9 a^2}\;\;\;\;and\;\;\;\;q^* =
-\frac{b}{3 a}
\end{equation}

Now the nonlinearity of the potential around $q=0$ can be
estimated from the general expression (\ref{29}) and (\ref{85}).
Thus we have

\begin{eqnarray}
V_N & = & V_2(q) + V_3 (q)\label{93}\\
& = & \sum_{n\geq3} \frac{1}{n!}\;V^{(n)}(q)\mid_{q=0}\;q^n +
\sum_{n\geq2} \frac{B_n}{n!}\;V_2^{(n)}(q)\nonumber\\
& = & -a\;q^3 - 3B_2\;a\;q - B_3\; a\label{94}
\end{eqnarray}

where $B_2$ and $B_3$ defined in (\ref{26}), are the quantum
corrections to the potential due to nonlinearity. The details of
the evaluation of $B_n$ are given in Sec.V. Linearizing $V_N$
around $q=-q_0$ we obtain the nonlinear part of the potential.

\begin{equation}\label{95}
V_N(q) \approx (3\;a\;q_0)\; q^2 + (3\;a\;q_0^2 - 3\;a\;B_2)\; q +
(a\;q_0^3 - a\;B_3)
\end{equation}

and its derivative

\begin{equation}\label{96}
V_N^{\prime}(q) = (6\;a\;q_0)\;q + (3\;a\;q_0^2 - 3\;a\;B_2)
\end{equation}

We are now in a position to write down the equation motion
(\ref{59}) for the unstable mode explicitly for the potential
concerned for the present problem

\begin{equation}\label{97}
\ddot{\rho} - \lambda_b^2 \;\rho = - \left[(4\;b\;u_{00}^2)
\;\rho + \left(\frac{4 b^2}{3 a}\;u_{00}^2 -
3\;a\;B_2\right)\right]
\end{equation}

or in the following form

\begin{equation}\label{98}
\ddot{\rho} + \lambda_0^2 \;\rho = r
\end{equation}

\begin{equation}\label{99}
where\;\;\;\; \lambda_0^2 = 4 b\; u_{00}^2 - \lambda_b^2
\end{equation}

\begin{equation}\label{100}
and\;\;\;\;r = - \left(\frac{4 b^2}{3 a}\;u_{00}^2 -
3\;a\;B_2\right)
\end{equation}

If the unstable mode start moving at $-\rho^*$ with total energy
$V^\ddag$, then the initial value of the momentum for the
unstable normal mode is

\begin{equation}\label{101}
\dot\rho(t=0) = - \frac{1}{u_{00}}\;\left(\frac{b\;\lambda_b}{3 a
}\right)
\end{equation}

The corresponding unstable co-ordinate is

\begin{equation}\label{102}
\rho(t=0) = - \frac{1}{u_{00}}\;\left(\frac{b}{3a}\right)
\end{equation}

For a round trip, the time elapsed is $t_p$; we have

\begin{eqnarray}
\rho(t=t_p) & = & - \frac{1}{u_{00}}\;\left(\frac{b}{3a}\right)\label{103}\\
and\;\;\;\;\;\dot\rho(t=t_p) & = &
\frac{1}{u_{00}}\;\left(\frac{b\;\lambda_b}{3 a
}\right)\label{104}
\end{eqnarray}

The equation for the unstable mode (\ref{97}) can be solved to
obtain

\begin{equation}\label{105}
\rho(t) = M\; \cos(\lambda_0 t) + N\; \sin(\lambda_0 t) +
\frac{r}{\lambda_0^2}
\end{equation}

where $M$ and $N$ are given by

\begin{equation}\label{106}
M = - \left(\frac{b}{3
u_{00}\;a}+\frac{r}{\lambda_0^2}\right),\;\;\;\;N = -
\frac{1}{\lambda_0 u_{00}}\;\left(\frac{b\;\lambda_b}{3 a }\right)
\end{equation}

The round trip time $t_p$ can be calculated form Eq.(\ref{105})
and its derivative equation for $\dot\rho$ and applying the
conditions (\ref{103}) and (\ref{104}) on them so that we have

\begin{equation}\label{107}
\cos(\lambda_0 t_p) =
\frac{M^2-N^2}{M^2+N^2}\;,\;\;\;\;\;\sin(\lambda_0 t_p) =
\frac{2\;M\;N}{M^2+N^2}
\end{equation}

Here we also note the range of $t_p$ as $\pi<\lambda_0 t_p<2\pi$.
An important point to emphasize is that $t_p$ contains a quantum
correction.

In order to calculate the energy loss due to the unstable mode we
now return to Eq.(\ref{66}) so that we write it as

\begin{equation}\label{108}
\Delta E = \frac{1}{2}\; \int_0^{t_p} dt \int_0^{t_p} dt^\prime\;
K_c (t-t^\prime)\; \zeta(t)\; \zeta(t^\prime)
\end{equation}

where $K_c(t)$ has the form given by Eq.(\ref{44}) since
$g_i^2=(u_{i0}/u_{00})^2$

Since $\widetilde{\gamma}(s)$ is given by (\ref{89}), $K_c(t)$ can
be calculated explicitly using (\ref{45}) as in the classical
theory \cite{pgh} to obtain

\begin{eqnarray}
K_c(t) = &\frac{1}{2}& \exp(- \xi t) \left[ (1 + 2 \epsilon) \cosh
(\sigma t) + \frac{(1 + 2 \epsilon)\xi-\lambda_b}{\sigma}\;\sinh
(\sigma t) \right]\nonumber\\ &-& \frac{1}{2}\;\exp(- \lambda_b
t)\label{110}
\end{eqnarray}

where $\xi$ and $\sigma$ are defined as
$\xi=\frac{1}{2}\;(\lambda_b+1/(\gamma\tau_c))$ and
$\sigma^2=\xi^2-\omega_b^2/(\gamma\tau_c\lambda_b)$. $\zeta(t)$ is
given by (\ref{61}) and using (\ref{96}) we obtain explicitly

\begin{equation}\label{111}
\zeta(t) = - 4\;b\;u_{00}^2 \;\rho(t) - \left( \frac{4 b^2}{3 a}
u_{00}^2 - 3 a B_2 \right)
\end{equation}

We then make use of the solution (\ref{105}) for $\rho(t)$ in
(\ref{111}) to obtain

\begin{equation}\label{112}
\zeta(t) = M_1 \cos(\lambda_0 t) + N_1 \sin(\lambda_0 t) - P
\end{equation}

where

\begin{eqnarray}
M_1 & = & - 4 M b u_{00}^2\nonumber\\
N_1 & = & - 4 N b u_{00}^2\label{113}\\
P & = & \frac{4 \;b \;u_{00}^2 \;r }{\lambda_0^2} + \frac{4 b^2
u_{00}^2}{3 a} - 3 a B_2 \nonumber
\end{eqnarray}

We are now in a position to calculate the energy loss $\Delta E$
of the unstable mode using (\ref{108}) from (\ref{110}) and
(\ref{112}). After a little bit of algebra we obtain

\begin{eqnarray}
\Delta E
&=&\frac{1+2\epsilon}{8}\;\left[R(\xi+\sigma)+R(\xi-\sigma)
\right]\nonumber\\ &-& \frac{(1+2\epsilon)\xi-\lambda_b}{8
\sigma}\left[R(\xi+\sigma)-R(\xi-\sigma)\right]-\frac{1}{4}\;R(\lambda_b)\label{114}
\end{eqnarray}

where

\begin{eqnarray}
R(z) = \int_0^{t_p} dt \int_0^{t_p} dt^\prime
e^{-z(t-t^\prime)}&&\left[M_1 \cos(\lambda_0 t) + N_1
\sin(\lambda_0 t) - P\right]\nonumber\\&\times& \left[M_1
\cos(\lambda_0 t^\prime) + N_1 \sin(\lambda_0 t^\prime) -
P\right]\label{115}
\end{eqnarray}

Explicit evolution of the integrals in (\ref{115}) yields

\begin{eqnarray}
R(z) & = & \left\{ \left[ \frac{P}{z}-\frac{M_1 z-N_1
\lambda_0}{\lambda_0^2+z^2} \right] \left[ \frac{P}{z} -
\left(\frac{M_1 z+N_1
\lambda_0}{\lambda_0^2+z^2}\right)\cos(\lambda_0 t_p)
\right]\right.\nonumber\\
 & + & \left.\left(\frac{M_1P\lambda_0+M_1N_1z-PN_1z}{z(\lambda_0^2+z^2)} -
 \frac{\lambda_0 z (M_1^2+N_1^2)}{(\lambda_0^2+z^2)^2}\right)\sin(\lambda_0
t_p ) \right\}e^{-z t_p}\nonumber\\
& + & \left[ \frac{P^2}{z} +
\frac{z(M_1^2+N_1^2)}{2(\lambda_0^2+z^2)}
\right]t_p + \frac{2P(M_1z-N_1\lambda_0)}{z(\lambda_0^2+z^2)} \nonumber\\
& - &
\frac{M_1^2z^2-N_1^2\lambda_0^2}{(\lambda_0^2+z^2)^2}+\frac{N_1z(M_1-4P)}{\lambda_0(\lambda_0^2+z^2)}
-\frac{P^2}{z^2}\label{116}
\end{eqnarray}

Our next task is to calculate the width $D$ of the distribution
function (\ref{71}). $D$ is defined in (\ref{69}) (and (\ref{68}))
and we now rewrite $\langle\delta E^2\rangle(=D)$ as

\begin{equation}\label{117}
\langle\delta E^2\rangle =\int_0^{t_p} dt \int_0^{t_p} dt^\prime\;
K_Q (t-t^\prime)\; \zeta(t)\; \zeta(t^\prime)
\end{equation}

where

\begin{equation}\label{118}
K_Q(t) = \sum_{i=1}^N g_i^2\; \hbar \lambda_i
\left(\overline{n}_i+\frac{1}{2}\right)\cos(\lambda_i t)
\end{equation}

We also recall Eq.(\ref{44})

\begin{equation}\label{119}
K_c(t)=\sum_{i=1}^N g_i^2\; \cos(\lambda_i t)
\end{equation}

By going over to continuum modes $K_c(t)$ and $K_Q(t)$ can be
expressed as

\begin{equation}\label{120}
K_Q(t)=\int_0^\infty d\lambda \;g^2(\lambda) f(\lambda) J(\lambda)
\cos(\lambda t)
\end{equation}

and

\begin{equation}\label{121}
K_c(t)=\int_0^\infty d\lambda \;g^2(\lambda) J(\lambda)
\cos(\lambda t)
\end{equation}

where $f(\lambda)=\frac{1}{2}\; \hbar \lambda \coth(\hbar
\lambda/2 k_{B} T)$. It can be easily shown that the above two
quantities can be related in the Fourier transform domain as

\begin{equation}\label{122}
\widetilde{K}_Q(\lambda)=f(\lambda) \widetilde{K}_c(\lambda)
\end{equation}

Here $\widetilde{K}_Q(\lambda)$ and $\widetilde{K}_c(\lambda)$ are
the cosine transform of $K_Q(t)$ and $K_c(t)$ respectively.

\begin{equation}\label{123}
\widetilde{K}_c(\lambda)=\left(\frac{2}{\pi}\right)^{1/2}\int_0^\infty
dt\; K_c(t)\;\cos(\lambda t)
\end{equation}

Making use of the expression $K_c(t)$ from Eq.(\ref{110}) in
Eq.(\ref{123}), followed by a multiplication of $f(\lambda)$,
yields $\widetilde{K}_Q(\lambda)$. The inverse transform of
$\widetilde{K}_Q(\lambda)$ finally gives

\begin{eqnarray}
K_Q(t) & = & \left(\frac{2}{\pi}\right)^{1/2}\int_0^\infty
d\lambda\;\frac{1}{2} \hbar \lambda \coth\left(\frac{\hbar
\lambda}{2 k_BT}\right) \; \cos(\lambda t)\nonumber\\
& \times & \left[\frac{W_1}{(\xi+\sigma)^2+\lambda^2}+
\frac{W_2}{(\xi-\sigma)^2+\lambda^2}-\frac{W_3}{\lambda_b^2+\lambda^2}\right]\label{124}
\end{eqnarray}

Here

\begin{eqnarray}
W_1 & = & \left(\frac{2}{\pi}\right)^{1/2}\left\{
\frac{1+2\epsilon}{4}-\frac{(1+2\epsilon)\xi-\lambda_b}{4\sigma}\right\}(\xi+\sigma)\nonumber\\
W_2 & = & \left(\frac{2}{\pi}\right)^{1/2}\left\{
\frac{1+2\epsilon}{4}+\frac{(1+2\epsilon)\xi-\lambda_b}{4\sigma}\right\}(\xi-\sigma)\label{125}\\
W_3 & = & \left(\frac{2}{\pi}\right)^{1/2}
\frac{\lambda_b}{2}\nonumber
\end{eqnarray}

with (\ref{112}) and (\ref{124}) we obtain the expression for
$\langle\delta E^2\rangle$ as follows.

\begin{eqnarray}
\langle\delta E^2\rangle & = & D\nonumber\\
 & = & \left(\frac{1}{2
\pi}\right)^{1/2}\hbar \int_0^\infty d\lambda\; \lambda
\coth\left(\frac{\hbar \lambda}{2 k_BT}\right)\nonumber\\
&\times&\left[\frac{W_1}{(\xi+\sigma)^2+\lambda^2}+
\frac{W_2}{(\xi-\sigma)^2+\lambda^2}-\frac{W_3}{\lambda_b^2+\lambda^2}\right]\nonumber\\
& \times &\int_0^{t_p}dt\int_0^{t_p}dt^\prime
\cos[\lambda(t-t^\prime)]\zeta(t)\zeta(t^\prime)\label{126}
\end{eqnarray}

The quantum expression of the depopulation factor in
Eq.(\ref{76}) primarily depends on the average quantum energy
loss $\Delta E$ and dispersion $\langle\delta E^2\rangle$. The
expressions of these two quantities have been given in
Eq.(\ref{114}) and Eq.(\ref{126}) for a cubic oscillator
(\ref{85}). An explicit evaluation of these quantities for
calculation of the rate or depopulation factor requires numerical
calculation. The parameter space chosen for illustration of the
main results depicted in Figs.2-5 is based on the earlier work by
Pollak \textit{et al} \cite{pgh}, Rips and Pollak \cite{rip},
Straub \textit{et al} \cite{s1} for a comparative study. The
results are given below.

In Fig.2 we exhibit the quantum turnover by plotting the quantum
rate, $\Gamma/\Gamma_Q$, where $\Gamma_Q=(\omega_0/{2 \pi})\chi
\exp \left[ -\; \frac{V^\ddag}{\hbar \lambda_0
(\overline{n}_0(\lambda_0) + \frac{1}{2})} \right]$, as a function
of dissipation parameter $\gamma$ at three different temperatures
$k_{B}T=5.0$ (dotted line) and $k_{B}T=1.0$ (dashed line) and
$k_{B}T=0.0$ (solid line) for a model potential with
$V^\ddag=20.0$, $a=0.03042$, $b=0.5$. The calculation here been
carried out with a non-Markovian friction kernel for
$\tau_c=1.333$ and the quantum correction has been taken upto
second order. In order to allow ourselves a fair comparison with
classical non-Markovian theory of Pollak, Grabert and H\"{a}nggi,
we compare the quantum (solid line) curves with the corresponding
classical (dotted line) curves at two different temperatures
$k_{B}T=1.0$ and $k_{B}T=2.0$ in Fig.3 for the same set of
parameter values as in Fig.2. The following two points are
noteworthy . We observe (Fig.2) in conformity with the earlier
observation \cite{h1,bk2} that as the temperature is lowered the
maximum at which the quantum turnover occurs shifts to the left
and the damping region that corresponds to classical energy
diffusion regime becomes exponentially small as one approaches to
absolute zero. Second, we observe that the strong friction tends
to make the dynamics more classical since the quantum correction
is suppressed by dissipation in this regime. On the other hand
differential behaviour of the rate in the classical and quantum
regime is felt at weak friction regime. This is in confirmation
with earlier observations. In Figs.4 and 5 we show a comparison
of the quantum rate, $\Gamma/\Gamma_Q$, calculated on the basis
of the present theory (solid line) with that of Rips and Pollak
(dotted line) for $V^\ddag=26.188$, $a=0.0266$ and $b=0.5$ at two
different scaled temperatures $k_{B}T=1.788$ and $k_{B}T=2.617$
respectively in the Markovian limit. The agreement is found to be
quite satisfactory. Thus although the depopulation factor in the
rate expression (\ref{76}) apparently differs in form from that of
Pollak and Rips based on quantum transition probability obtained
from the solution of master equation, the good numerical
agreement between them can not be overlooked (since the remaining
factor in $\Gamma/\Gamma_Q$, the Grote-Hynes factor is same in
both the expressions). Figs.4 and 5 therefore serve as a
consistency check of the present calculation.

\section{{\bf Summary and Conclusion}}

Based on a quantum Langevin equation we have constructed a
c-number Hamiltonian for a system plus N-oscillator bath model.
This allows us to formulate a normal mode analysis to realize a
c-number version of the multidimensional transition state theory
and to derive a quantum expression for the total decay rate of
metastable state. The result is valid for arbitrary damping
strength and noise correlation and temperature down to vacuum
limit. The theory is illustrated on a cubic potential with
non-Ohmic dissipation. The following pertinent points are
noteworthy.

(i) We have shown that the  expression for quantum rate
coefficient is a product of five terms , \textit{e.g.} , classical
well frequency, Kramers'-Grote-Hynes factor, a vacuum corrected
or generalized Wolynes factor representing quantum transmission
and reflection, an exponential term corresponding to Wigner
canonical thermal distribution \textit{i.e.} the generalized
Arrhenius term and a quantum depopulation factor. Of these the
Wigner term and the vacuum corrected Wolynes term refer to
equilibriation of quantum particle in the well and therefore
corresponds to a c-number multidimensional transition state
result. Since the distribution unlike the Boltzmann is valid even
as $T\rightarrow0$, the quantum effect due to heat bath can be
well accounted by these terms even below the activated tunneling
regime.

(ii) The quantum depopulation factor has a form which is very much
similar to its classical counterpart, although intrinsically the
relevant quantities determining this factor, \textit{i.e.} , the
average quantum energy loss and dispersion are quantum mechanical
in their content.

(iii) The classical limit of the quantum rate expression depends
on Wigner, generalized Wolynes and depopulation factors. It is
easy to see that they reduce to Arrhenius factor, unity and
classical depopulation factor, respectively in the limit $\hbar
\omega \ll k_BT$.

(iv) Since the quantization of the nonlinear system mode adds an
additional contribution to depopulation factor arising out of the
weak coupling between the unstable mode and the stable modes the
rate is significantly modified.

(v) The theory takes into account of the quantum effects in full
due to heat bath, while the quantum correction due to
nonlinearity of the system potential can be calculated
systematically order by order to a good degree of accuracy. In the
illustrative example with cubic potential we have considered the
quantum correction of the second order.

(vi) The quantum depopulation factor interpolates the energy
diffusion limited regime to spatial diffusion regime and depicts
the correct turnover scenario down to absolute zero. The maximum
of the turnover shifts to the low damping regime as one approaches
the absolute zero. The quantum rate is significantly enhanced
over the classical rate in the weak damping regime while strong
dissipation overwhelms the quantum correction.

(vii) An important aspect of the present theory is that it takes
care of activation and tunneling within a unified description and
is equipped to deal with the rate at temperature down to vacuum
limit. This is a distinct advantage over path integral Monte
Carlo method since numerically the relevant propagator poses
serious problem as the temperature approaches absolute zero.

(viii) The present theory is a synthesis of the classical
formalism of normal mode analysis and weak coupling theory of
unstable and stable modes for calculation of average energy loss
and dispersion within the framework of our recently developed
c-number quantum Langevin equation. It is important to emphasize a
distinct advantage of the present treatment. The present c-number
formalism allows us to formulate the Guassian transition
probability that the energy of the unstable mode changes for one
period of motion within a "classical-like" prescription, rather
than by solving a quantum master equation restricted to a
Markovian description.

The quantum turnover theory as presented here is based on a
canonical quantization procedure and positive definite Wigner's
thermal distribution for harmonic oscillators rather than path
integral or master equation formalism. It can be readily applied
to other models , \textit{e.g.} , a double well oscillator and to
the problems of dephasing and related issues. The systematic
improvement can be made by taking care of the quantum corrections
of higher orders.
\newpage

{\bf Acknowledgement}\\
We are thankful to Dr. B. C. Bag for discussions. The authors are
indebted to the Council of Scientific and Industrial Research for
partial financial support under Grant No. 01/(1740)/02/EMR-II.

\newpage

\begin{center}
{\bf Figure Captions}
\end{center}

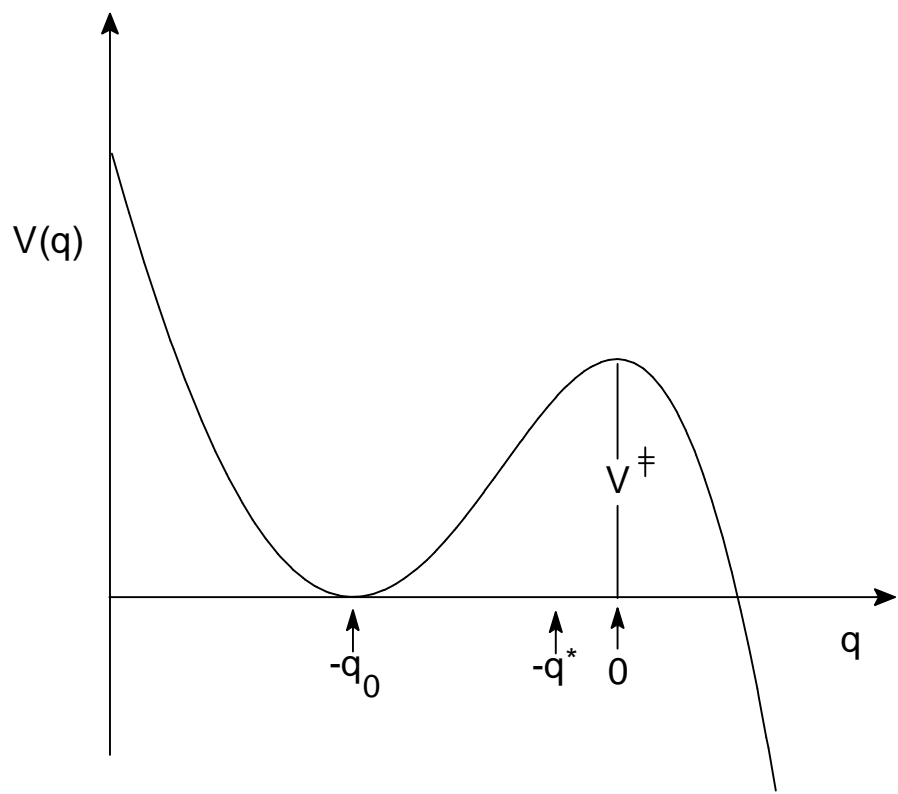
Fig.1. A schematic plot of the potential defined in Eq.(6.1)

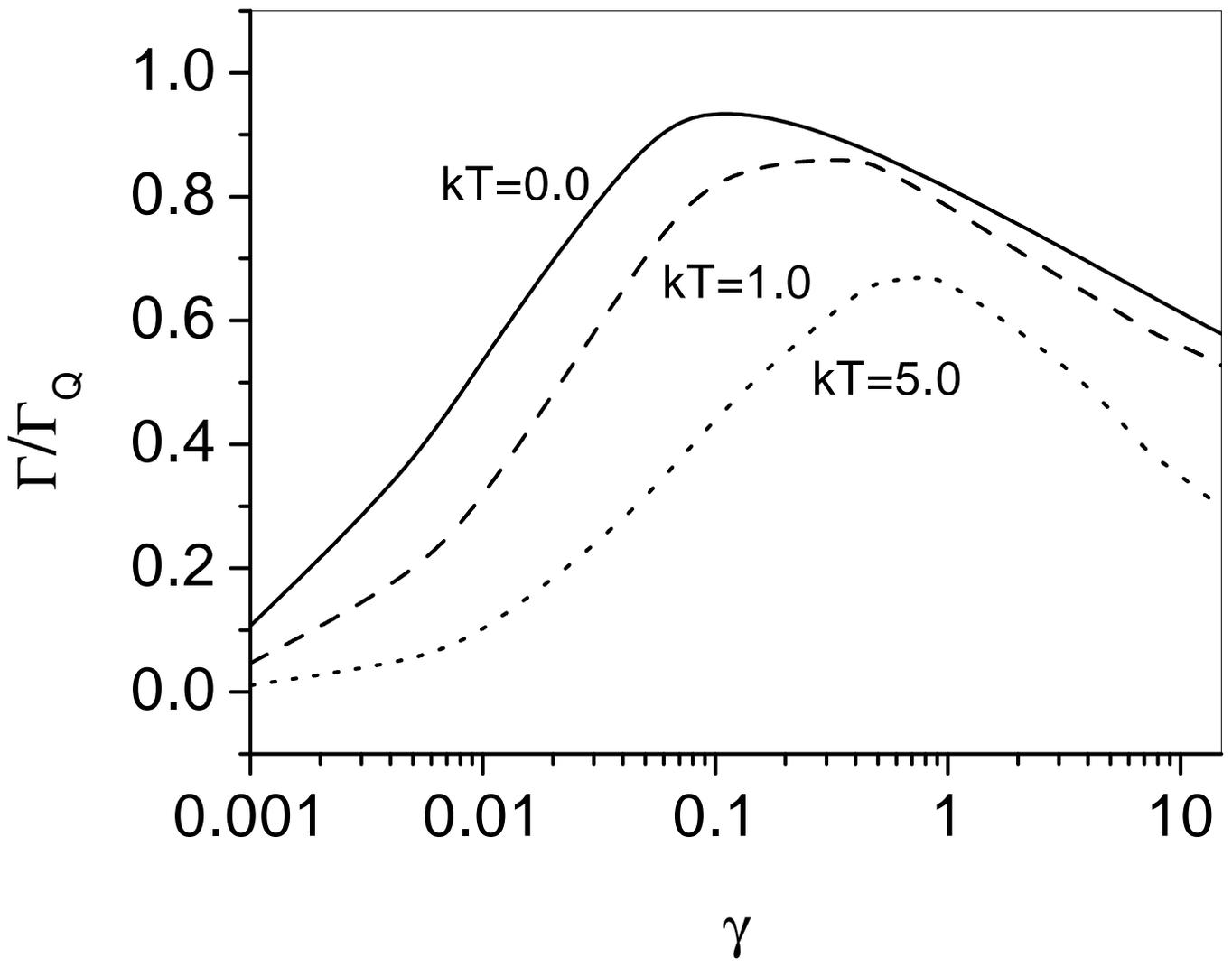
Fig.2. Quantum rate, $\Gamma/\Gamma_Q$, (where
$\Gamma_Q=(\omega_0/{2 \pi}) \chi \exp \left[ -\;
\frac{V^\ddag}{\hbar \lambda_0 (\overline{n}_0(\lambda_0) +
\frac{1}{2})} \right]$ ), in the non-Markovian regime is plotted
against dissipation parameter, $\gamma$, for three different
temperatures $k_{B}T=0.0$ (solid line), $k_{B}T=1.0$ (dashed
line) and $k_{B}T=5.0$ (dotted line) with $V^\ddag=20.0$,
$a=0.03042$, $b=0.5$ and $\tau_c=1.333$.

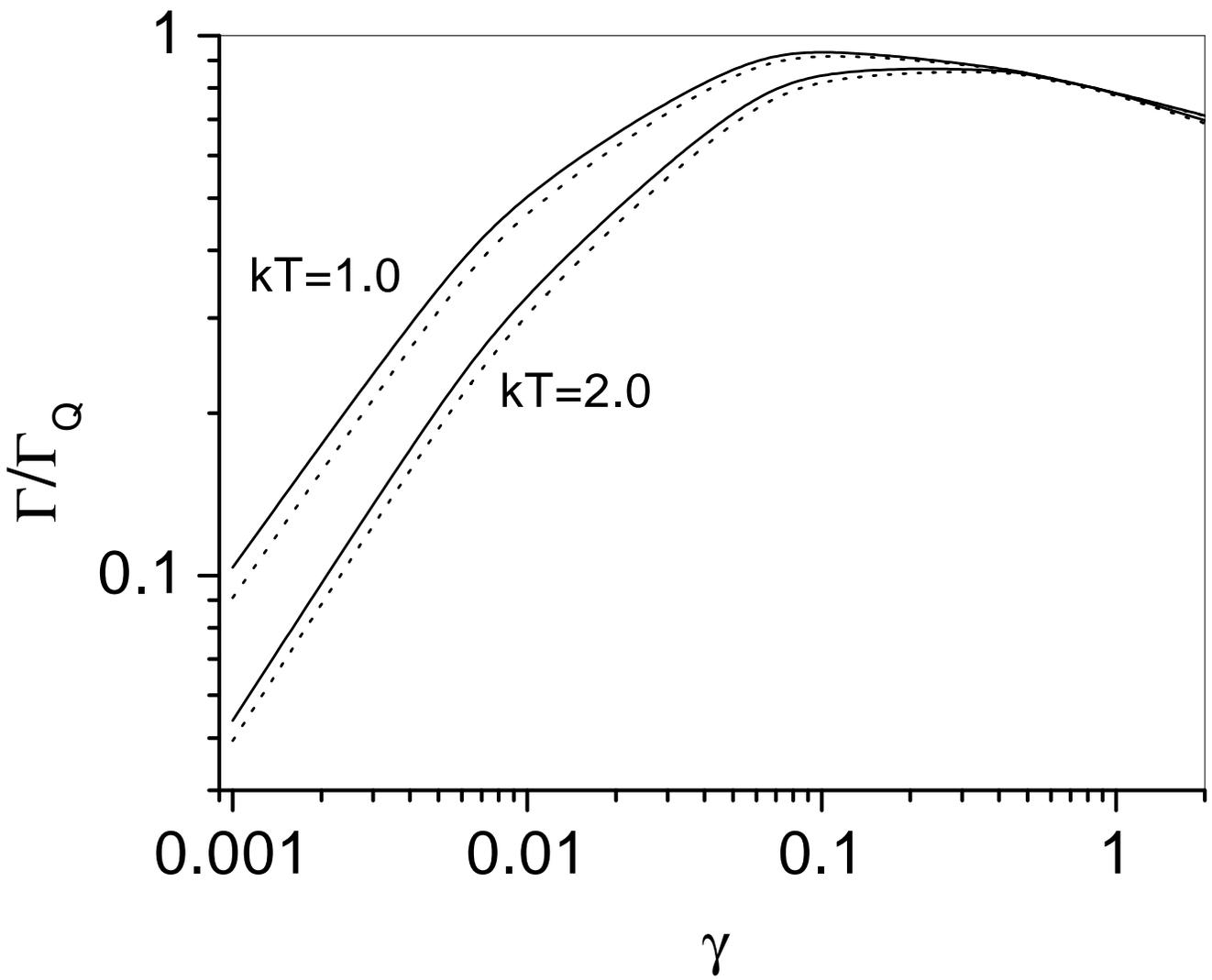
Fig.3. Quantum turnover (solid line) is compared with classical
turnover \cite{pgh} of Pollak, Grabert and H\"{a}nggi (dotted
line) in the non-Markovian regime for two different temperatures
$k_{B}T=1.0$ and $k_{B}T=2.0$, other parameters remain same as
Fig.2.

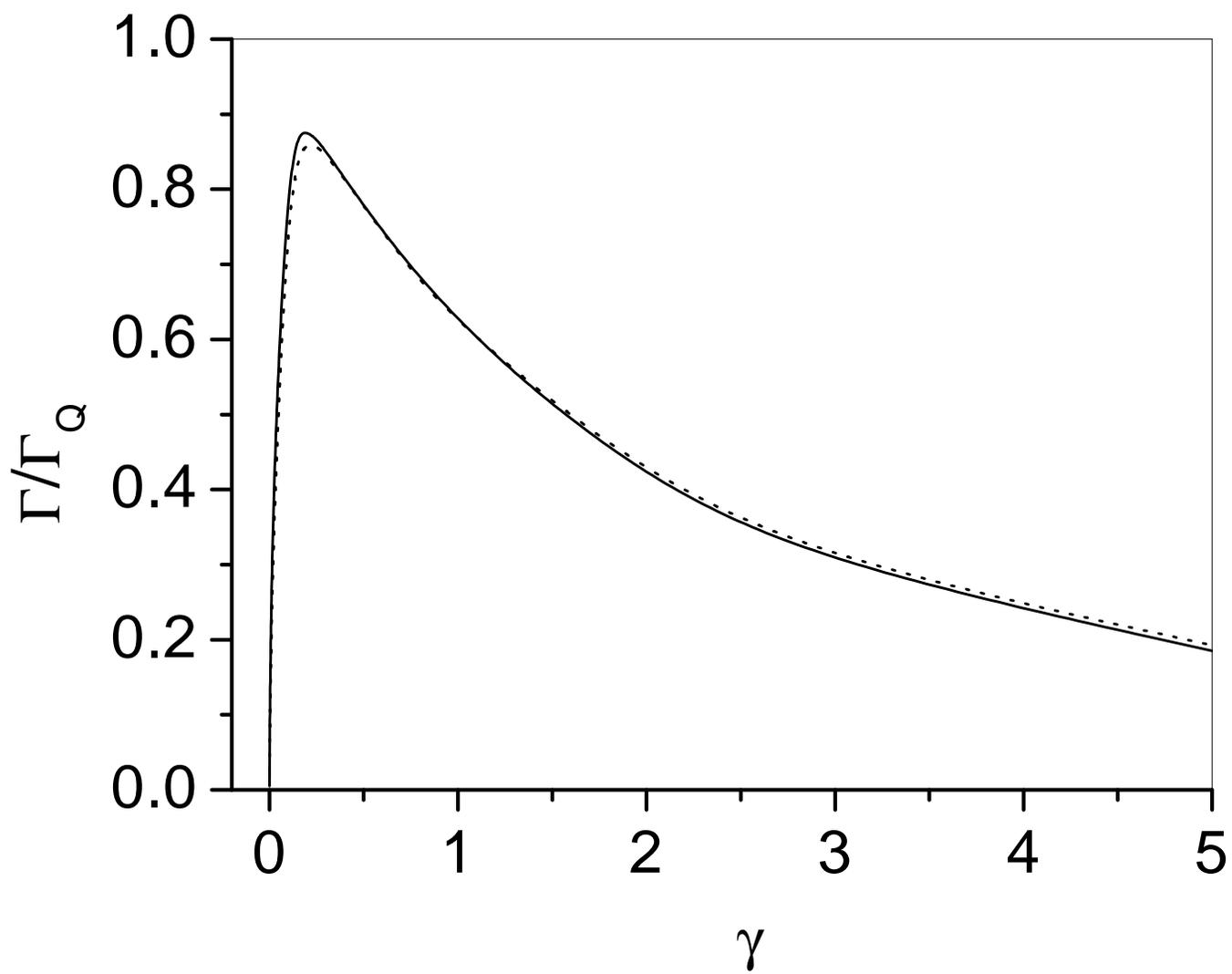
Fig.4. The comparison of Kramers' turnover for $V^{\ddag}=26.188$,
$a=0.0266$, $b=0.5$ at $kT=1.788$. The solid line represents the
quantum turnover by the present method, the dotted line
represents the calculation by Rips and Pollak \cite{rip}.

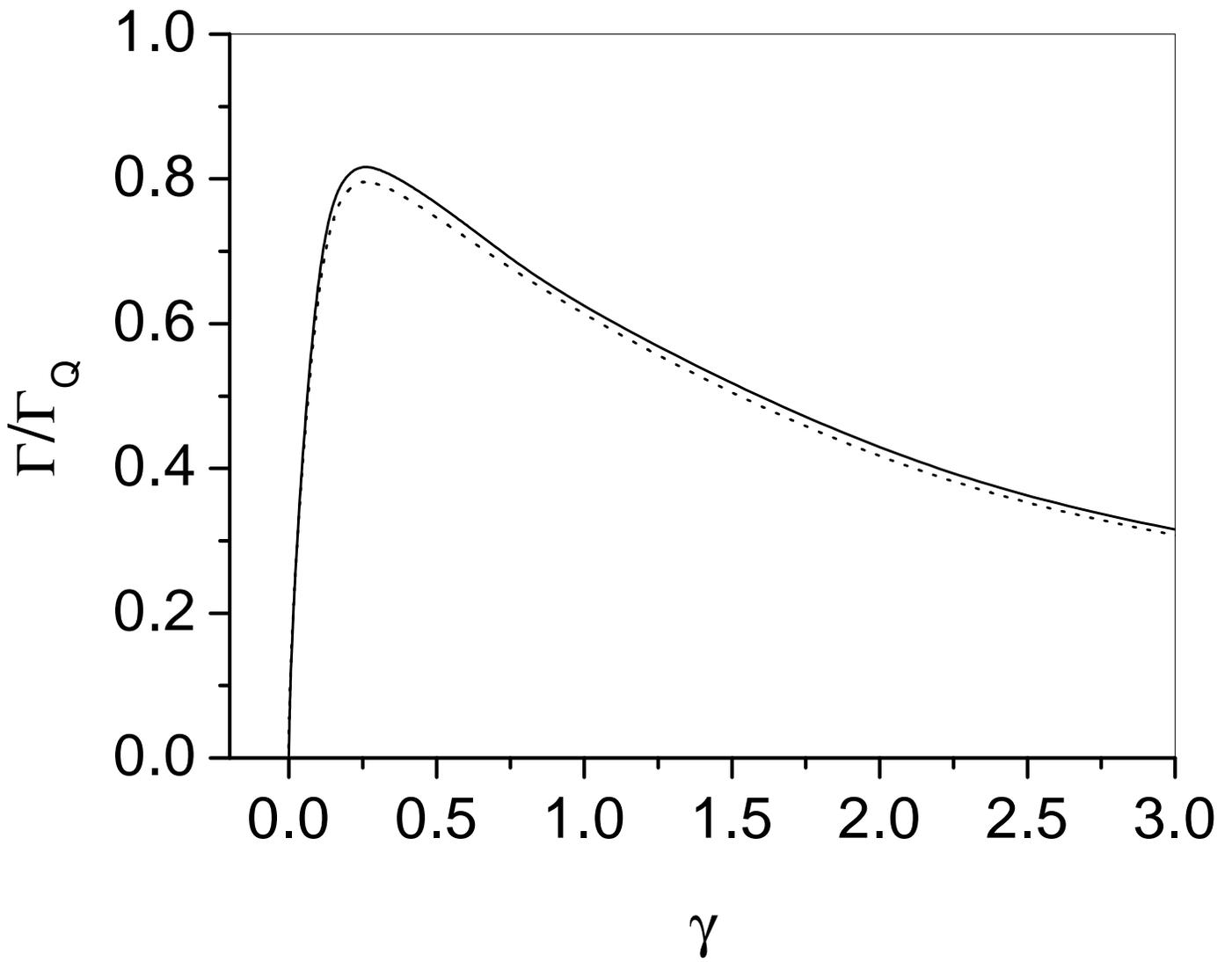
Fig.5. The same as in Fig.4 but for $k_B T=2.617$

\end{document}